\documentclass[12pt,letterpaper]{article}          %% LaTeX 2e (preferred)
\usepackage{osajnl}
\usepackage[draft]{hyperref} % optional
\usepackage{setspace}

\begin{document}
%%%%%%%%%%%%%%%%%%%%%%%%%%%%%%%%%%%%
\doublespacing

\title{Microscopic surface structure of C/SiC composite mirrors
for space cryogenic telescopes}

\author{Keigo Enya}

\address{
Department of Infrared Astrophysics,
Institute of Space and Astronautical Science,
Japan Aerospace Exploration Agency,
Yoshinodai 3-1-1, Sagamihara, Kanagawa 229-8510, Japan
}

\email{enya@ir.isas.jaxa.jp}

%--------------------------------

\author{Takao Nakagawa}

\address{
Department of Infrared Astrophysics,
Institute of Space and Astronautical Science,
Japan Aerospace Exploration Agency,
Yoshinodai 3-1-1, Sagamihara, Kanagawa 229-8510, Japan
}

\email{nakagawa@ir.isas.jaxa.jp}

%--------------------------------

\author{Hidehiro Kaneda}

\address{
Department of Infrared Astrophysics,
Institute of Space and Astronautical Science,
Japan Aerospace Exploration Agency,
Yoshinodai 3-1-1, Sagamihara, Kanagawa 229-8510, Japan
}

\email{kaneda@ir.isas.jaxa.jp}

%--------------------------------

\author{Takashi Onaka}

\address{
Department of Astronomy, 
Graduate School of Science, University of Tokyo,
Hongo 7-3-1, Bunkyo-ku, Tokyo 113-0033, Japan
}

\email{onaka@astron.s.u-tokyo.ac.jp}

%--------------------------------

\author{Tuyoshi Ozaki}

\address{
Advanced Technology R \& D Center, Mitubishi Electric Corporation,
Miyashimo 1-5-57, Sagamihara, Kanagawa 229-1195, Japan
}

\email{Ozaki.Tsuyoshi@wrc.melco.co.jp}

%--------------------------------

\author{Masami Kume}

\address{
Advanced Technology R \& D Center, Mitubishi Electric Corporation,
Miyashimo 1-5-57, Sagamihara, Kanagawa 229-1195, Japan
}

\email{Kume.Masami@wrc.melco.co.jp}

%--------------------------------

\begin{abstract}

We report on the microscopic surface structure of carbon-fiber
reinforced silicon carbide (C/SiC) composite mirrors that have 
been improved for Space Infrared telescope for Cosmology 
and Astrophysics (SPICA) and other cooled telescopes. 
The C/SiC composite consists of carbon-fiber, 
silicon carbide and residual silicon. 
Specific microscopic structures are found on the surface of
the bare C/SiC mirrors after polishing. 
These structures are considered to be caused by the different 
hardness of those materials. 
The roughness obtained for the bare mirrors is 
20\,nm rms for flat surfaces and 100\,nm rms for curved surfaces. 
It was confirmed that a SiSiC slurry coating is effective 
in reducing the roughness down to 2\,nm rms. 
The scattering properties of the mirrors
were measured at room temperature and also at 95\,K. 
No significant change was found in the scattering 
properties through cooling, which suggests that the microscopic 
surface structure is stable with changes in temperature down 
to cryogenic values. 
The C/SiC mirror with the SiSiC slurry coating is a 
promising candidate for the SPICA telescope.

\end{abstract}

%\ocis{000.0000, 999.9999.}% REPLACE WITH CORRECT OCIS CODES FOR YOUR ARTICLE
%                          % NOTE: \ocis{} IS ALIASED TO \pacs{} BUT MUST
%                          % FORMAT THE TERMS CORRECTLY FOR EACH JOURNAL

\maketitle %% NULL FUNCTION WITH LATEX 2e; required for REVTeX4
%%%%%%%%%%%%%%%%%%%%%%%%%%%%%
\section{Introduction}\label{sec01}
%%%%%%%%%%%%%%%%%%%%%%%%%%%%%

The development of light-weight mirrors is one of key technologies
required to realize the next-generation  astronomical infrared space
telescopes, for which mirrors much larger than the present ones are
needed{\cite{nakagawa2004}}$^,${\cite{onaka2005a}}.
Silicon carbide (SiC) and its variants are promising
candidates of the material for light-weight mirrors.  
The major characteristics of SiC are its high strength, 
thermal conductivity, and especially the high ratio of 
Young's modulus to density. 
In particular, because of its high stiffness, mirrors using
these materials can be made with thin ribs and skin structures, so
that the weight can be much reduced.

SiC mirrors have been adopted for the AKARI 
mission{\cite{murakami2004}} and the
Herschel Space Observatory{\cite{pilbratt2004}}
(ASTRO-F is a Japanese infrared astronomical satellite 
which was launched in February in 2006 and is now called AKARI). 
The telescope on AKARI has a 710-mm-diameter primary
mirror consisting of porous and chemical 
vapor deposited SiC{\cite{kaneda2005}$^,$ \cite{kaneda2003}}. 
The entire AKARI telescope system is cooled by 
vapor from liquid helium to 6\,K. 
The Herschel Space Observatory is an infrared astronomical
mission from the European Space Agency with a telescope equipped
a 3.5\,m diameter sintered SiC mirror{\cite{pilbratt2004}}. 
This satellite is to be launched in 2008 and operated 
with a passive cooling system{\cite{sein2003}}.
NASA has urged the development of the James Webb Space 
Telescope (JWST), which will
have deployable light-weight mirrors of beryllium for a 6.5\,m
aperture telescope in space{\cite{aabelhaus2004}}.

Space Infrared telescope for Cosmology and Astrophysics
(SPICA) is a next generation mission led by Japan 
for infrared astronomy planned to
be launched into a halo orbit around one of the Sun-Earth libration
points (L2) in the 2010s{\cite{nakagawa2004}$^,$ \cite{onaka2005a}}. 
The SPICA telescope is designed to have a 3.5 m diameter aperture, 
and will be cooled to 4.5\,K by radiative
cooling and mechanical cryo-coolers for observations at infrared
wavelengths{\cite{onaka2005b}}. 
It will be optimized for observations in the 5-200\,$\mu$m 
wavelength, and have image quality of diffraction limit  
at 5\,$\mu$m.
The requirement for the surface roughness of the mirrors 
for the SPICA telescope is 20 nm rms or better at 4.5\,K. 
Table \ref{table01} summarizes the optical specifications 
of the SPICA telescope.

Carbon-fiber reinforced silicon carbide (C/SiC) composite is
one of the promising materials for the SPICA 
telescope{\cite{ozaki2004}},
while the other is sintered SiC. C/SiC composite has
characteristics similar to SiC and can provide a higher
fracture toughness than SiC{\cite{ozaki2004}}. 
Furthermore, the C/SiC
composite has another advantage in which  it can be machined
easily during the C/C stage (see below), and therefore, a
lightweight rib structure and complex shaped design for the
support mechanism can be realized.

A sufficiently small surface roughness is one of the most
fundamental properties required for mirror
applications{\cite{enya2004}}. 
However, it is not guaranteed that a surface
with sufficiently small roughness can be realized with the
C/SiC mirror because of the complicated nature of composite
materials. 
SiSiC slurry coating has often been applied to
improve roughness, although it requires additional processes
in the fabrication{\cite{ozaki2004}}. 
Furthermore, the cryogenic performance
of C/SiC and SiSiC slurry coatings has not yet been well
investigated. The coefficients of thermal expansion (CTE) of
the SiC{\cite{cte_sic}}, Si{\cite{cte_si}}, 
and carbon-fiber{\cite{cte_cf}} are 
2.6\,$\times$\,10$^{-6}$\,K$^{-1}$, 
3.3\,$\times$\,10$^{-6}$\,K$^{-1}$, 
and 0.8-1.1\,$\times$\,10$^{-6}$ K$^{-1}$
at room temperature,
respectively. The difference in the CTEs of these materials
causes thermal stress in the C/SiC composite when the mirrors
are cooled to cryogenic temperatures, and consequently the
microscopic structure might change when cooled.

In this paper, we report on measurements and the improvement
in the microscopic surface structures of both bare and slurry
coated mirrors made of the newly developed C/SiC composite. We
also carried out measurements of the scattering properties of
these mirrors at  $\sim$\,300\,K and 95\,K to examine possible 
changes in the surface roughness caused by cooling.

%%%%%%%%%%%%%%%%%%%%%%%%%%%%%
\section{Surface of the C/SiC composite mirror and its measurement}
%%%%%%%%%%%%%%%%%%%%%%%%%%%%%

\subsection{Manufacturing process of the mirror}{\label{subsec_2A}}

The fabrication process of the improved C/SiC composite 
is summarized briefly below;  the details are described 
in Ozaki et al. 2004{\cite{ozaki2004}}.

\begin{enumerate}

\item

Preparation of the carbon-fiber preform : for the new C/SiC
composite, the preform is composed of pitch-based milled fibers and
binder.

\item
The CF preform is impregnated with coal tar pitch. 
They are carbonized and graphitized in an inert
atmosphere to produce the carbon-fiber carbon matrix (C/C).

\item
The C/C material is machined into the end-product geometry. 
The C/C material is easily milled 
into light-weight structures with thin ribs and skins.

\item
If necessary, C/C segments can be joined by adhesives 
to build larger or more complicated structures

\item
The C/C substrate is infiltrated with liquid silicon to 
react with the carbon matrix and form SiC. During this process, 
the joints between the segments become SiC.

\item
 If necessary, a SiSiC slurry coating is applied.

\end{enumerate}

As a result of improvements in the manufacturing process,
particularly owing to the increase in the volume fraction of
pitch-based carbon-fiber to more than 30\,\% 
and to the adequate control
of the reactivity, several material properties have been significantly
improved. For example, the bending strength, Young's modulus, the
toughness, and the thermal conductivity become 
200\,MPa, 320\,GPa, 3.9\,MPa\,m$^{1/2}$, 
and 160\,W\,m$^{-1}$\,K$^{-1}$
at room temperature respectively, 
while the conventional values are 
160\,MPa, 260\,GPa, 2.4\,MPa\,m$^{1/2}$,  
and 125\,W\,m$^{-1}$\,K$^{-1}$.

\subsection{Measurement of the surface properties}

\begin{center}
\begin{figure}
\begin{center}
\includegraphics[width=12cm]{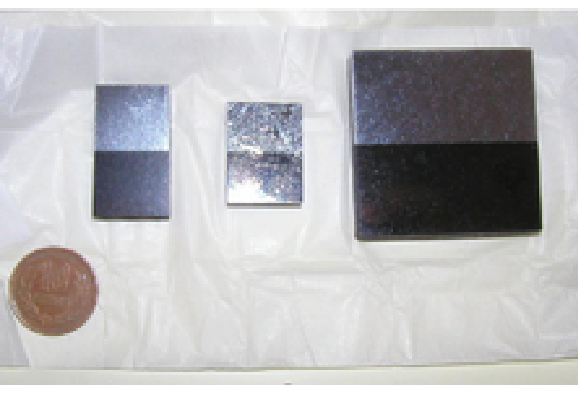}
\caption{Enya et al.
}
\label{fig01}
\end{center}
\end{figure}
\end{center}

We prepared small test mirrors to investigate the surface 
properties and provide quick feedback to the fabrication process. 
A photograph of the test mirrors is shown in figure {\ref{fig01}}. 
They are flat mirrors produced by employing different fabrication 
processes for the C/SiC composite. 
The right square mirror in figure {\ref{fig01}}
is made using the improved C/SiC.  
Half of the mirror surface is coated with aluminium
for the light scattering measurements 
described in section {\ref{sec04}}.

Microscope interferometers are used to observe the 
surface structure of the test mirrors. The interferometers provide
high-precision three-dimensional measurements, in which the resolution
is sub-micron parallel to the mirror surface, and sub-nm in the
vertical direction. Results of the three dimensional measurements are
highly useful not only for estimating the roughness, but also for
understanding the origin of the roughness and for improving the
surface quality. For each measurement, we used one of the available
interferometers, either the ZYGO Maxim-NT, the WYKO NT1100 in National
Astronomical Observatory Japan, or the ZYGO New-View in
RIKEN. 
It is 
difficult to apply  microscope interferometers for 
mirrors at cryogenic temperatures,
so measurements of the scattering properties
were made as described in section {\ref{sec04}}.

The top-left panel of figure {\ref{fig02}} shows a typical result of the
measurement by the microscope interferometer ZYGO Maxim-NT for the
surface of a conventional C/SiC composite mirror shown in the left of
figure {\ref{fig01}}. 
The top-right of figure {\ref{fig02}} is a photographic image obtained
through the viewer of the interferometer with the fringe pattern
removed. The light gray region in the photograph corresponds to
residual silicon. The gray region indicates SiC and slender
carbon-fiber. The SiC regions tend to be $\sim 100$ nm higher than the
silicon regions. This tendency is consistent with the difference
between the hardness of the SiC and Si; the Mohs hardness of the SiC
is 9 and that of Si is 7, respectively{\cite{hardness}}. 
The carbon-fiber regions tend to be 
trenches with $\sim 100$ nm depth. 
The average roughness of the whole
surface is 40\,nm rms.

The results of these measurements were taken 
into account in the material production process 
improvement trials{\cite{ozaki2004}} and were very
convenient because of their immediacy. 
The middle panel of figure {\ref{fig02}}
shows the data of the improved C/SiC composite mirror obtained in the
same manner as the first one. The height difference between the SiC
and the residual silicon regions has become smaller as a result of the
reduction in the amount of residual silicon in the improved
material and polishing improved for the new composite. 
Carbon-fiber is still the dominant factor in the surface
roughness. Surface roughness to be 20\,nm rms 
has been achieved
for flat mirrors made using the improved C/SiC composite.

%%%%%%%%%%%%%%%%%%%%%%%%%%%%%
\section{Microscopic surface quality of the spherical mirrors}
%%%%%%%%%%%%%%%%%%%%%%%%%%%%%

\subsection{Surface without  SiSiC slurry coating}

To test the surface quality of a more realistic mirror 
than the test pieces, we fabricated a 160\,mm diameter 
spherical test mirror with a flat outer region near the edge
using the improved C/SiC composite material, as shown 
in figure {\ref{fig03}}. 
The radius of curvature of the spherical part is 900 mm. 
Both the spherical and flat regions were polished. 
Three segments were joined at three of the six 
ribs to examine the effects of the joints.

An example of the microscopic surface structure of the flat region
is shown in the top left and right of figure {\ref{fig04}}. 
The roughness of the
outer flat region is similar to that of the improved flat test mirrors
made from the improved C/SiC composite ($\sim$\,20\,nm rms), 
which satisfies
the requirement of the SPICA telescope.

However, the structure of the spherical part was different from
the outer flat region. The middle-left and right 
of figure {\ref{fig04}} show the
results of measurements on the spherical part. The roughness of the
spherical part is typically 100\,nm rms, which is significantly worse
than that of the flat region. Mirrors of this quality can be useful
for mid- to far-infrared observations, however, SPICA requires a
better quality. The major sources of the roughness are the trenches
due to the carbon-fiber and pits due to the residual silicon. As a
result, we conclude that the current bare mirrors made of C/SiC
composite do not meet the requirements for the SPICA telescope.

\subsection{Surface of the joint  of segments}

The joint is one of the key techniques needed to realize large
monolithic mirrors from small segments. The joint material for joining
C/SiC composite parts together is made of SiC. The CTE, Young's
modulus, and the other properties of the joint are similar to those of
the mirror.

The bottom of figure {\ref{fig04}} shows 
the microscopic surface structure of
the joint of the 160\,mm spherical mirror. The width of the joint in
this mirror is about 0.5\,mm. Most of the area of the joint consists of
SiC, but there is also residual silicon here. We find that the height
of the residual silicon region tends to be a few tens of nanometers
lower than the SiC region, similar to the difference in height between
the SiC and silicon regions in the flat test mirrors. Based on this
result, we adjusted the viscosity of the adhesive to optimize it for
the improved C/SiC composite. Finally, the thickness of the joint now
becomes 0.2\,mm and the residual silicon is much reduced.

\subsection{Surface with SiSiC slurry coating}

Since the presence of carbon-fiber is a major factor in making the
polishing process difficult and increasing the surface roughness, we
applied a SiSiC slurry coating{\cite{ozaki2004}}$^,${\cite{enya2004}}
to the C/SiC to make a surface
without carbon-fiber. The typical thickness of the slurry coating is
200\,$\mu$m.

The polished surface of the improved C/SiC composite mirror
with the SiSiC slurry coating appears quite different from
those of bare polished C/SiC composite mirrors. The bottom of
figure {\ref{fig02}} shows the microscopic 
surface structures of a slurry
coated and polished flat mirror measured with the ZYGO
New-View. The typical roughness is $\sim$\,2\,nm rms or less. 
The improvement in the roughness is highly significant and the
obtained surface may be applied even for visible light
observations. In the photograph (bottom-right of the figure
{\ref{fig02}}), 
the dark gray spot-like regions correspond to SiC, and
the light gray region indicates silicon. We examined the
correlation between the 3-D data and the optical appearance,
which indicates that the SiC region has become bumpy. This
trend is similar to that seen in the bare polished surface of
the improved C/SiC composite. However, both the scale and
height of the SiC regions are much smaller in the slurry
coated surface than in the bare ones.

We prepared a 160\,mm diameter spherical test mirror with a
SiSiC slurry coating similar to that in the previous
section. The microscopic surface structure of this mirror was
measured with an interferometer. The roughness and appearance
of the surface are similar to those of the flat test mirror,
not only in the outer flat region but also in the central
spherical part. The surface of the joint was also examined and
no significant differences were found between the joint and
other regions.

Therefore, we conclude that the C/SiC composite mirror with
the SiSiC slurry coating is a promising candidate for the
SPICA telescope.

Table {\ref{table02}} 
summarizes the microscopic surface properties of the
mirrors developed in this work.

%%%%%%%%%%%%%%%%%%%%%%%%%%%%%
\section{Light  scattering measurements}\label{sec04}
%%%%%%%%%%%%%%%%%%%%%%%%%%%%%

\subsection{Experimental method} 
%%%%%%%%%%%%%%%%%%%%%%%%%%%%%%%%%%%%

Space-borne infrared telescopes must be cooled to achieve high
sensitivity and therefore
the mirrors for the telescope must satisfy their 
specifications  at cryogenic temperatures. 
The C/SiC composite consists of materials with different
CTEs as described in section {\label{sec01}}, 
and consequently the surface of the
mirrors can be deformed or damaged on the microscopic scale by thermal
stress at low temperatures. Direct observations with a microscope
interferometer are difficult to perform at cryogenic temperatures. In
this section, we examine the surface roughness of C/SiC composite
mirrors using the light scattering method.

We carried out measurements of the scattered light from test
mirrors at $\sim$\,300\, K  and $95$\,K. 
The configuration used for the
measurement is shown in figure {\ref{fig05}}.
As described above, the
diameter of the carbon fiber is about 10\,$\mu$m and its length and
the size of the residual Si region are larger  than 10\,$\mu$m.  
Light scattered by these structure can spread into a few
degrees due to the diffraction of light at  0.63\,$\mu$m,
so we set up
an experiment to measure the scattered light in the range of a
few degrees. A flat test mirror was installed in a dewar with
a BK7 flat window. The test mirror was thermally connected to
the cold work surface of the dewar, which was cooled by liquid
nitrogen. A He-Ne laser of 0.63\,$\mu$m wavelength 
and 1\,mm diameter
was irradiated onto the aluminum coated part of the mirror.

The reflected light was measured by a Si photo-diode with an
electrical amplifier. No AC chopping was used. The profile of
the scattered light was measured by scanning the detector in a
linear direction. The background signals were measured without
the light irradiation and were subtracted from the
measurements. The incident angle of the irradiated beam was 
5\,degrees. Each test mirror was measured both at  
$\sim$\,300\,K and 95\,K. 
The temperature was measured with a platinum resistance
temperature sensor placed directly on the surface of the
mirror. As a reference, a BK7 flat mirror with an aluminum
coating was also measured.

\subsection{Results} 
%%%%%%%%%%%%%%%%%%%%%%%%%%%%%%%%%%

Obtained profiles of the scattered light from the mirrors
are shown in  figure {\ref{fig06}}. 
In the plot, the data are
transformed into the intensity in a solid angle.  
Profiles in the top figure indicate the results for 
the improved C/SiC composite mirror
without a SiSiC slurry coating
shown on the right of figure {\ref{fig01}}. 
Compared with the results for the BK7
mirror, significant scattered light is clearly detected. No
appreciable difference is found between the results 
at 300\,K and 95\,K. 
The results for the SiSiC coated mirror in the bottom
of figure {\ref{fig02}} are presented in the middle of
figure {\ref{fig06}}. 
No significant difference is seen between the SiSiC coated mirror and
the BK7 mirror surface. 
No appreciable difference is detected between
the results at $\sim$\,300\,K and 95\,K 
also for this test mirror. 
The limit to the sensitivity of the scattered 
light measurements is 10$^{-5.5}$\,$\sim$\,10$^{-6}$ 
for the scale of figure {\ref{fig06}}. 
The measured positions on the mirror
surface may change with cooling because of the shrinkage of the cold
work surface. However, the stability of the scattered light between
$\sim$\,300\,K and 95\,K indicates that this effect is not significant.

The sensitivity of the present measurement is limited by
the detector dark current and the laser light scattered at the window
and in the air.  
The former was estimated to be about $5\,\times\,10^{-6}$
in units of figure 6 when measured with the laser light off. 
This dark current has been subtracted from the plots of figure 6.  
The latter cannot be measured directly in the present setup,
but the comparison of the C/SiC with and without slurry coat
(and BK7) 
measurements indicates that the present measurements 
have an total uncertainty of $\sim\,\pm\,1\times\,10^{-6}$, 
which is limited by the uncertainty in the 
dark current.

The scale of the surface structures that produce the peak
profile in the central $\pm$\,0.2\,degree 
is estimated to be $\sim 180 \mu$m
if diffraction dominates the scattering. The length of the
trenches in the carbon-fiber regions are of a similar scale,
though it is difficult to distinguish the scattered light from
the reflected light itself in the vicinity of the specularly
reflected beam. On the other hand, the width of the trenches
is much smaller and the length of them is not uniform. The
width of the trenches and the length of the shortest trenches
are $\sim 10 \mu$m which corresponds to $\pm 3.6$ degree. 
The scattering
angle from smaller structures or structures of other geometry
is expected to be larger. Thus the results of the
measurements suggest that the roughness caused by the
carbon-fiber does not increase during cooling.

From the surface roughness of the C/SiC mirror obtained from
the interferometer measurements and the results of the light
scattering measurements, we can estimate the total integrated
scatter of the mirror at cryogenic temperatures at optical and
infrared wavelengths. Here we use the following estimate for
the total integrated scattering{\cite{davies1954}}:

\begin{equation}
TIS = \frac{E_s}{ E_a} = \left( \frac{4 \pi \sigma  \cos
\theta_i}{\lambda} \right)^2, {\label{eq01}}
\end{equation}
where $TIS$ is the total integrated scattering, 
$E_s$ is the scattered energy
and $E_a$ is the total energy of the incident beam, 
$\sigma$ is 
rms roughness,
$\theta_i$ is the incident angle, 
and $\lambda$  is the wavelength. The values
of $TIS$ derived from the roughness of the test 
mirrors shown in table {\ref{table02}}
are 0.17 and $1.7 \times 10^{-3}$ for the 
bare C/SiC flat mirror and the mirror
with the SiSiC slurry coating, respectively, 
with $\theta_i$\,=\,5\,degree 
and $\lambda$\,=\,0.63\,$\mu$m.
At $\lambda$\,=\,5\,$\mu$m,
$TIS$ derived from equation (\ref{eq01}) is 
$2.5 \times 10^{-3}$, $2.5 \times 10^{-5}$, 
0.06 for the bare C/SiC flat mirror, the mirror with the
SiSiC slurry coating, and the bare spherical mirror,
respectively. Therefore, the fraction of the energy lost by scattering
at the microscopic surface structure is estimated to be insignificant
in the mid-infrared region for slurry coated mirrors. To estimate the
$TIS$ from the scattering experiment directly, 
measurement of light with a larger scattering angle is needed.

The difference in the profiles between the bare mirror and the
others shown in figure {\ref{fig06}} 
is significant at which the relative
intensity is smaller than $\sim$ 10$^{-4.5}$.
With this value, scaled by
equation (\ref{eq01}), for the sensitivity of detecting scattered light
$\sigma$\,=\,20\,nm for the bare mirror, the limit to the sensitivity
for measuring roughness is $3.6 \sim 6.5$\,nm rms.
This limit is larger than the roughness of the BK7 mirror 
measured by the interferometer, 
and therefore it is reasonable that no significant 
scattered light was  detected with  the BK7 mirror.

The present experiment was carried out with the available He-Ne laser.  
The measurement sensitivity for roughness can be improved 
if we use a laser with shorter wavelegths.  
The measurement with infrared wavelengths is also useful
to confirm directly the scattering property for the SPICA 
application.  
Examinations of the scattering properties are complementary to
measurements with microscope interferometers. Measurements of
the scattered light profiles provide the optical performance
of mirrors directly. Furthermore, this method can be applied
at cryogenic temperatures. On the other hand, measurements by
interferometer give us useful information about the origin of
the roughness.
In this work, 
the results of the two method are consistent with each other

%%%%%%%%%%%%%%%%%%%%%%%%%%%%%
\section{Summary}
%%%%%%%%%%%%%%%%%%%%%%%%%%%%%

The microscopic surface structure of mirrors made of improved C/SiC
composite was investigated. The improved C/SiC is a candidate material
for the SPICA mirrors because of its superior properties: high
toughness, high stiffness, machinability, and the feasibility of
making large single dish mirrors.

We examined flat test mirrors and 160\,mm diameter spherical
mirrors with and without a SiSiC coating. The surface of the
bare C/SiC composite consists of carbon-fiber, silicon carbide
and silicon. 
Specific  structures were found at the surface of
the bare C/SiC surface after polishing, which is consistent
with the difference in hardness of those materials.  

The achieved surface roughness was 20\,nm rms or less for bare C/SiC
composite flat mirrors, which just about satisfies the requirement of
the SPICA telescope. For the curved surface of the bare C/SiC mirror,
the nominal roughness was about 100\,nm, which is larger than that
obtained for flat mirrors. At present, the roughness of the curved
bare surface is not small enough for the SPICA telescope, though such
mirrors still can be useful for mid to mid- and far-infrared
observations. 
The joint between the segments was examined 
and an improvement was made in
the adhesive process. The amount of residual silicon was reduced and
the width of the joint was narrowed down to 0.2\,mm from 0.5\,mm. 
We can
confirm that the SiSiC slurry coating is effective for reducing the
roughness of both flat and curved surfaces down to 2\,nm rms. For
polished slurry coated mirrors with a spherical surface, no
significant difference was detected between the joint and the other
parts.

The change in the scattering properties of the C/SiC composite
at cryogenic temperatures was examined with a 0.63\,$\mu$m
laser. Light scattered from the microscopic surface structures
was detected for bare C/SiC composite mirrors, while SiSiC
slurry coated mirrors showed scattering properties similar to
glass mirrors. No significant change was found between the
scattering properties at 95 K and  $\sim$\,300\,K, 
suggesting that the
microscopic structures of the surface are stable against
cooling. 
The $TIS$ of the examined mirrors at room temperature
and 95 K is discussed.

We conclude that the mirrors made with the improved C/SiC with
a SiSiC slurry coating satisfy the specifications for the
SPICA telescope, and therefore they are one of 
quite promising candidates
for the SPICA mirrors. Other candidate materials
(e.g. sintered SiC) are also being examined for the SPICA
telescope using the same methods. On the other hand, the
surface quality of the bare mirrors is suitable for use at
mid- and far- infrared wavelengths where cooling of the optics
is quite important.

%%%%%%%%%%%%%%%%%%%%%%%%%%%%%
\section*{Acknowledgment}
%%%%%%%%%%%%%%%%%%%%%%%%%%%%%

We are grateful to M. Otsubo and N. Ebizuka, who provided us with
opportunities to use the microscopic interferometers of their
institutes and useful comments. We also thank Florent Le Ne'chet for
providing support for our experiments. The SPICA project is managed by
the Institute of Space and Astronautical Science (ISAS), Japanese
Aerospace Exploration Agency (JAXA) in collaboration with universities
and institutes. We would like to thank all of the SPICA working group
members. This work was supported in part by a grant from the Japan
Science and Technology Agency (JST).

%%%%%%%%%%%%%%%%%%%%%%%%%%%%%%%%%%%%%%%%%%%%%%%%%%

%\appendix

\newpage

\vspace*{50mm}

\begin{center}
\begin{table}
\begin{center}
\caption{The optical specifications for the SPICA telescope}
\label{table01}
\begin{tabular}{lll}
\hline
\hline
  Parameter      &  specification &   \\
\hline
  Telescope optics type &  On axis Ritchey-Chretien    \\
                    &  Consists of monolithic mirrors\\
  Image quality     &  5 $\mu$m diffraction limit \\
  Diameter of the primary mirror           &  3.5 m          &  \\
  Optimal wavelength        & 5 - 200   $\mu$m\\
  Working temperature         & 4.5 K &       \\
  Weight                 & $<$  700 kg               &    \\
  Focal length           & $\sim$ 18m \\
  Field-of-view diameter & $\sim$ 30 arcmin \\
  Total wave-front error & $<$   350 nm (rms)  &     \\
  Surface roughness of the mirrors      & $<$ 20 nm (rms) &    \\
\hline
\end{tabular}
\end{center}
\end{table}
\end{center}

\newpage

%%%%%%%%%%%%%%%%%%%%%%%%%%%%%%%%%%%%%%%%%%%%

\vspace*{50mm}

\begin{center}
\begin{table}
\begin{center}
\caption{Microscopic surface quality of C/SiC composite mirrors}
\label{table02}
\begin{tabular}{lllcl}
\hline
\hline
  Sample  &    &    Coating   & Roughness (rms)  &  Origin  of roughness\\
\hline
  Flat test mirror: & Conventional   &  No    &  40 nm   &  Si dimple,  carbon-fiber trench\\
                  & Improved       &  No     &  20 nm   &  Carbon-fiber  trench \\
  $\phi 160$mm mirror: &  Flat part  &  No      & 20 nm    &  Carbon-fiber trench\\
                     & Spherical part  &  No    & 100 nm   &  Carbon-fiber trench\\
\hline
  Flat test mirror:   &            &  SiSiC slurry   & 2 nm   &  SiC bump  \\
    $\phi 160$mm  mirror: & Flat part  &  SiSiC  slurry  & 2 nm   &  SiC bump  \\
                      & Spherical part  &  SiSiC  slurry   & 2 nm   & SiC bump \\
 \hline
\end{tabular}
\end{center}
\end{table}
\end{center}

\newpage

\section*{List of figure captions}

\vspace{5mm}
\noindent Figure  1.
 Examples of flat test mirrors for examining the
manufacturing process of the C/SiC composite material, the polishing
technique and the microscopic surface structures. 
The left and the right mirror are sample before and after  
the trial in which the carbon fiber was more dispersed and
the residual Si was reduced.
The middle and the right mirror  are  test piece in the trial 
of the improvement.
The size of the right mirror is 
50\,mm\,$\times$\,50\,mm\,$\times$\,10\,mm.
Half of the polished surface is
coated with aluminum for measurements of the scattered light described
in section \ref{sec04}.

\vspace{5mm}
\noindent Figure  2.
Top: The left figure shows the microscopic surface
structures obtained from the interferometer (ZYGO Maxim-NT) for a flat
test mirror of the conventional C/SiC composite. 
The right figure is a photographic image of the same 
part of the surface with visible light. 
The relatively light gray area indicates residual Si. 
The dark gray part corresponds to SiC and carbon-fiber. 
The average surface roughness of this sample is 40\,nm rms. 
Middle: Microscopic surface structure of a flat test 
mirror of the improved C/SiC composite. 
The average value of surface roughness is 20\,nm rms. 
Bottom: Microscopic surface structure of a SiSiC slurry 
coated flat test mirror using the ZYGO New-View. 
The relatively bright and dark gray areas in the
photograph correspond to silicon and SiC, respectively. SiC regions
are higher than silicon regions. The morphology of the slurry coated
surface is quite different to that of the bare surface. The average
roughness of the coated surface is 2 nm rms. Note that the vertical
scale is not the same for each figure.

\vspace{5mm}
\noindent Figure  3.
The left and the right show front and back view of
the 160\,mm diameter test mirror of the improved C/SiC
composite. The radius of curvature of the spherical part is 900\,mm,
while the periphery is flat. 
Three segments 
which were fan shaped with 120 degree central angle
were joined at the ribs.

\begin{center}
\begin{figure}
\begin{center}
\includegraphics[height=6cm]{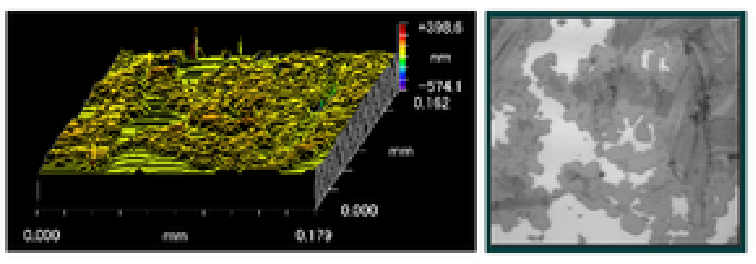}\\
\vspace{5mm}
\includegraphics[height=6.03cm]{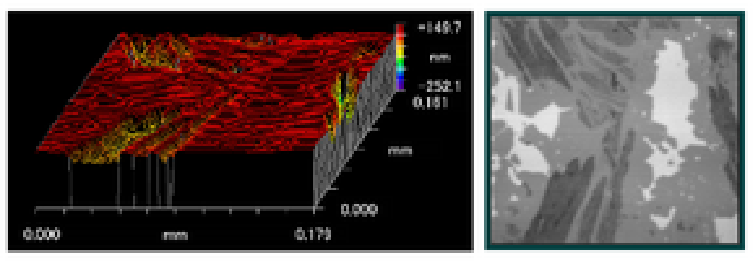}\\
\vspace{2mm}
\includegraphics[height=7.05cm]{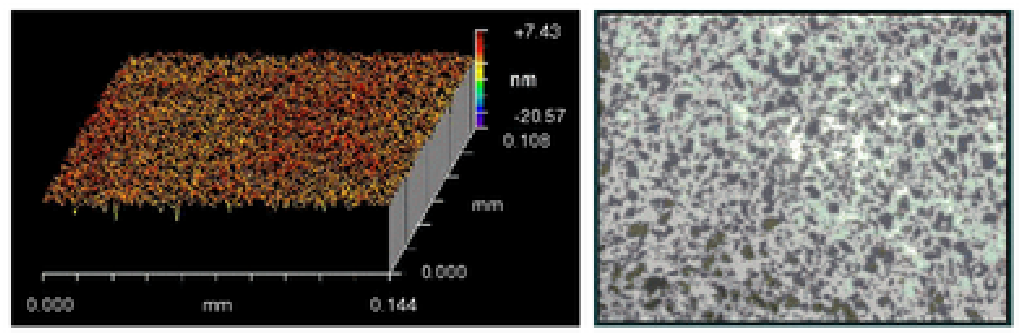}
\caption{Enya et al.}
\label{fig02}
\end{center}
\end{figure}
\end{center}

\begin{center}
\begin{figure}
\begin{center}
\includegraphics[height=6cm]{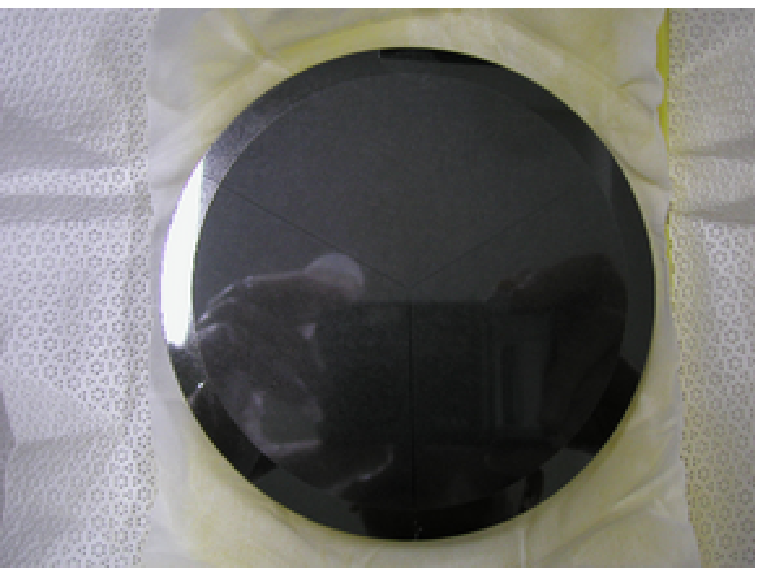}
\includegraphics[height=6cm]{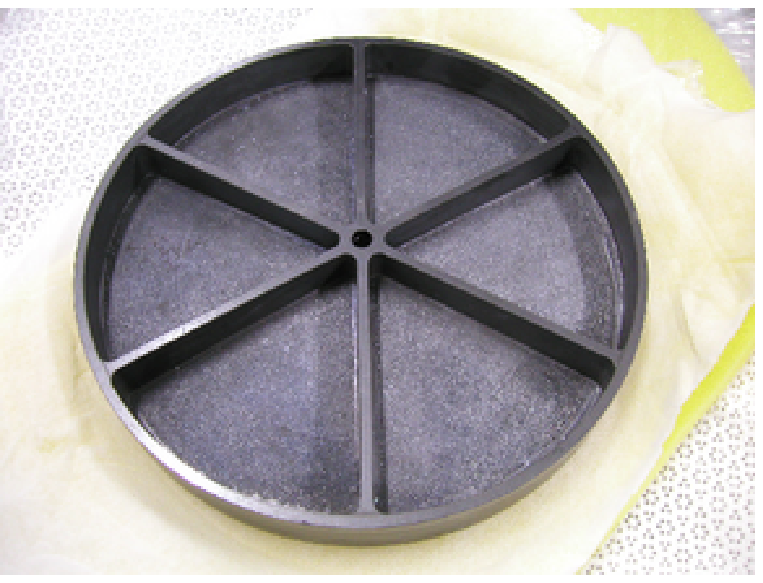}
\caption{Enya et al.}
\label{fig03}
\end{center}
\end{figure}
\end{center}

\begin{center}
\begin{figure}
\begin{center}
\includegraphics[height=14cm]{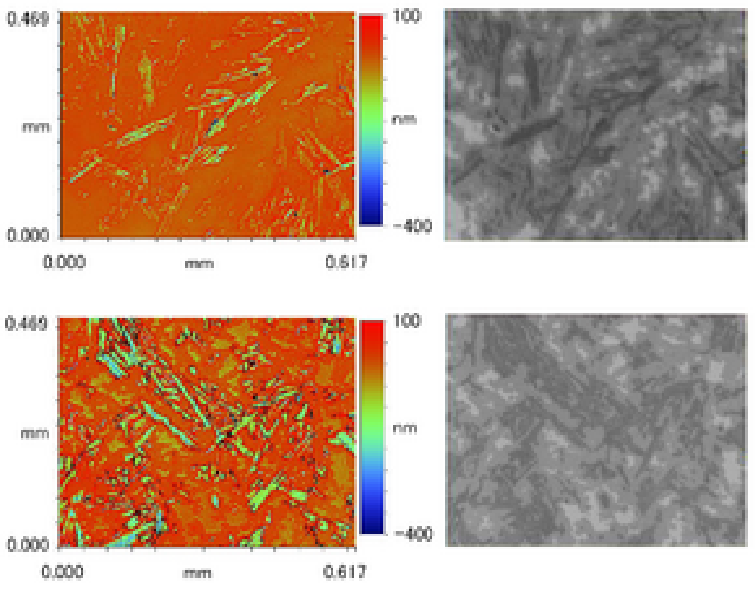}
\includegraphics[height=7.06cm]{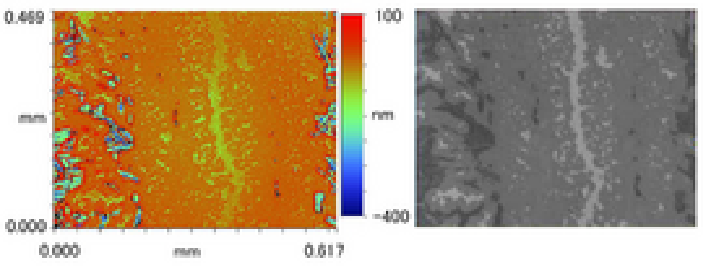}
\caption{Enya et al.}
\label{fig04}
\end{center}
\end{figure}
\end{center}

\begin{center}
\begin{figure}
\begin{center}
\includegraphics[height=8cm]{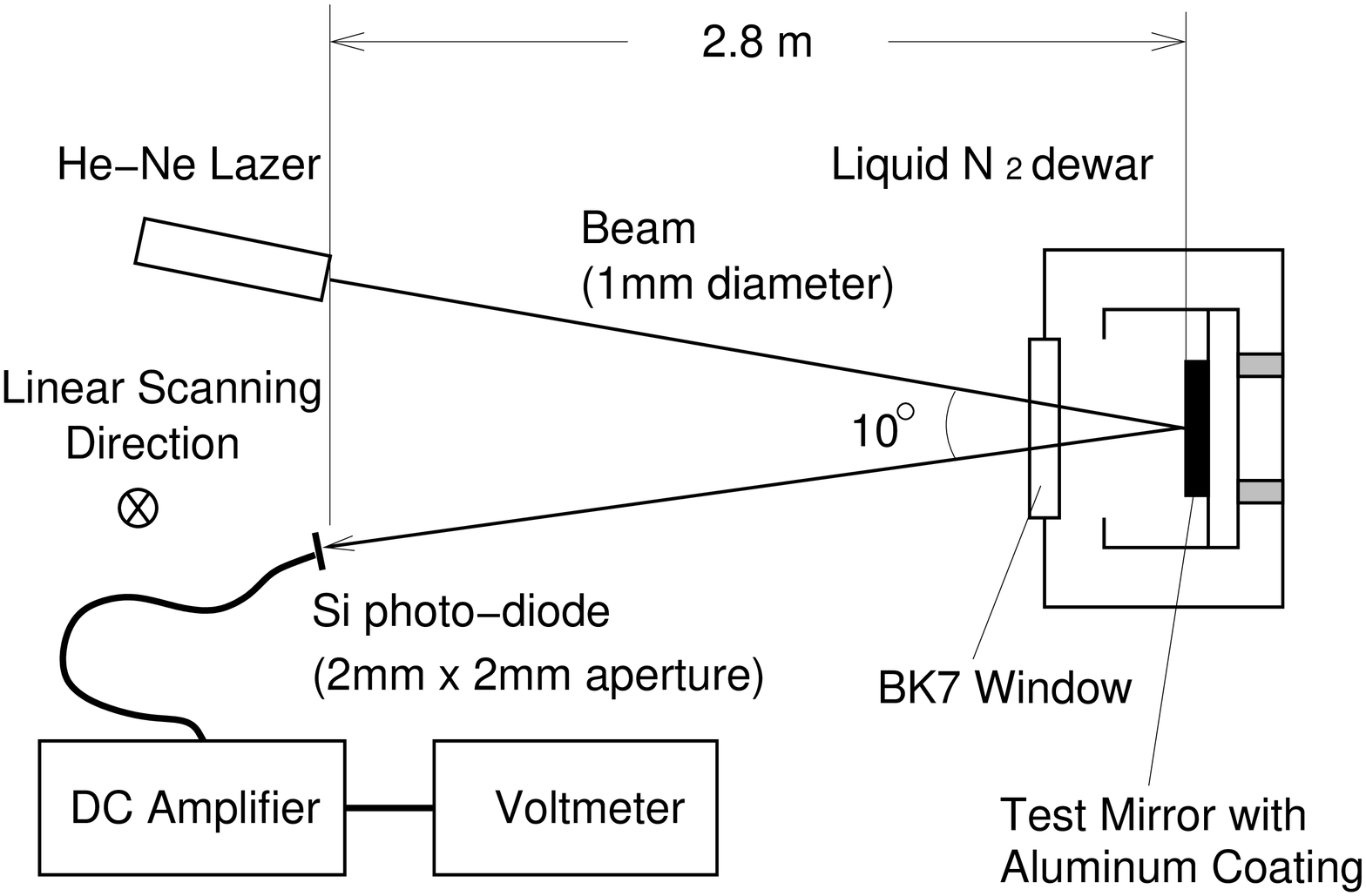}
\caption{Enya et al.}
\label{fig05}
\end{center}
\end{figure}
\end{center}

\begin{center}
\begin{figure}
\begin{center}
\includegraphics[width=10cm]{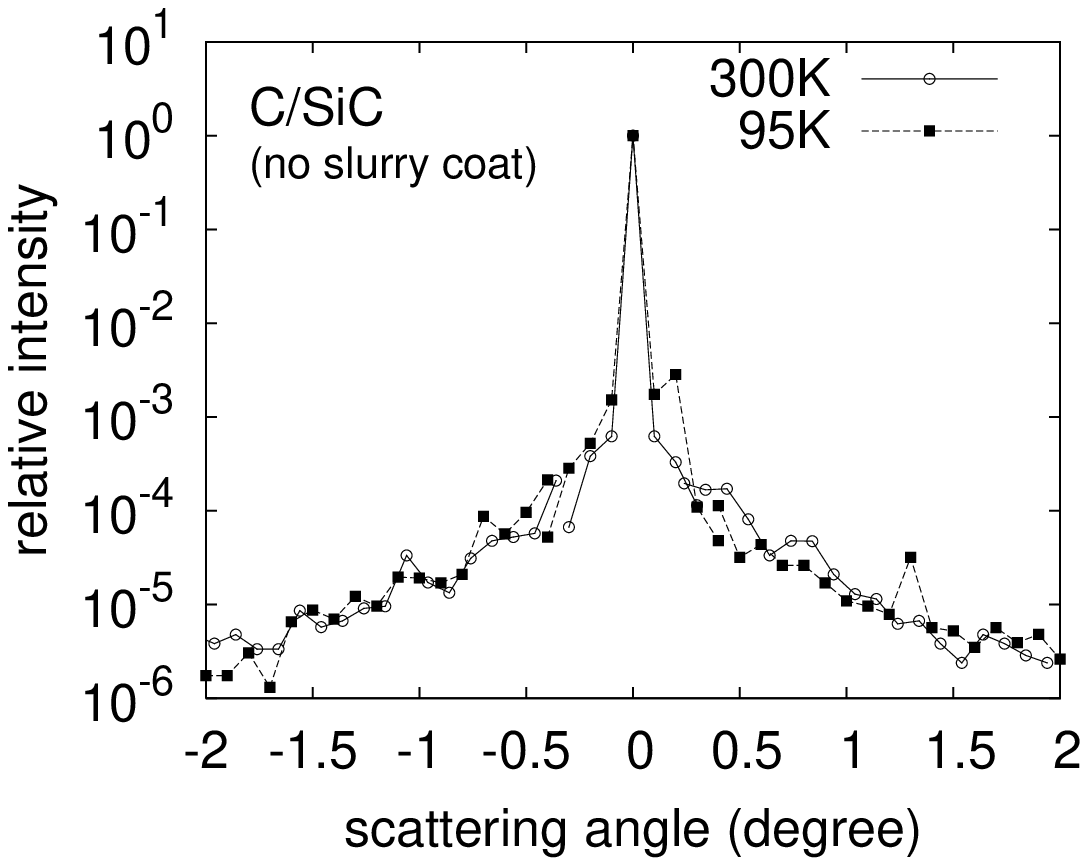}\\
\includegraphics[width=10cm]{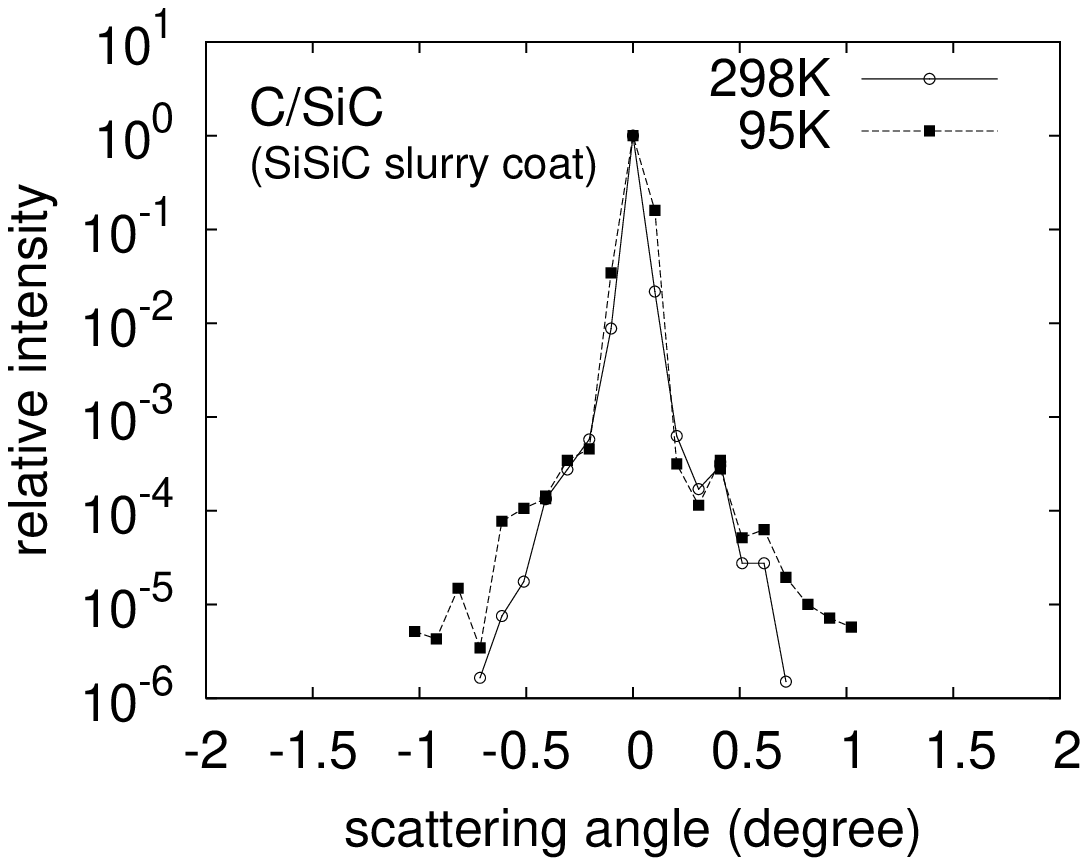}\\
\includegraphics[width=10cm]{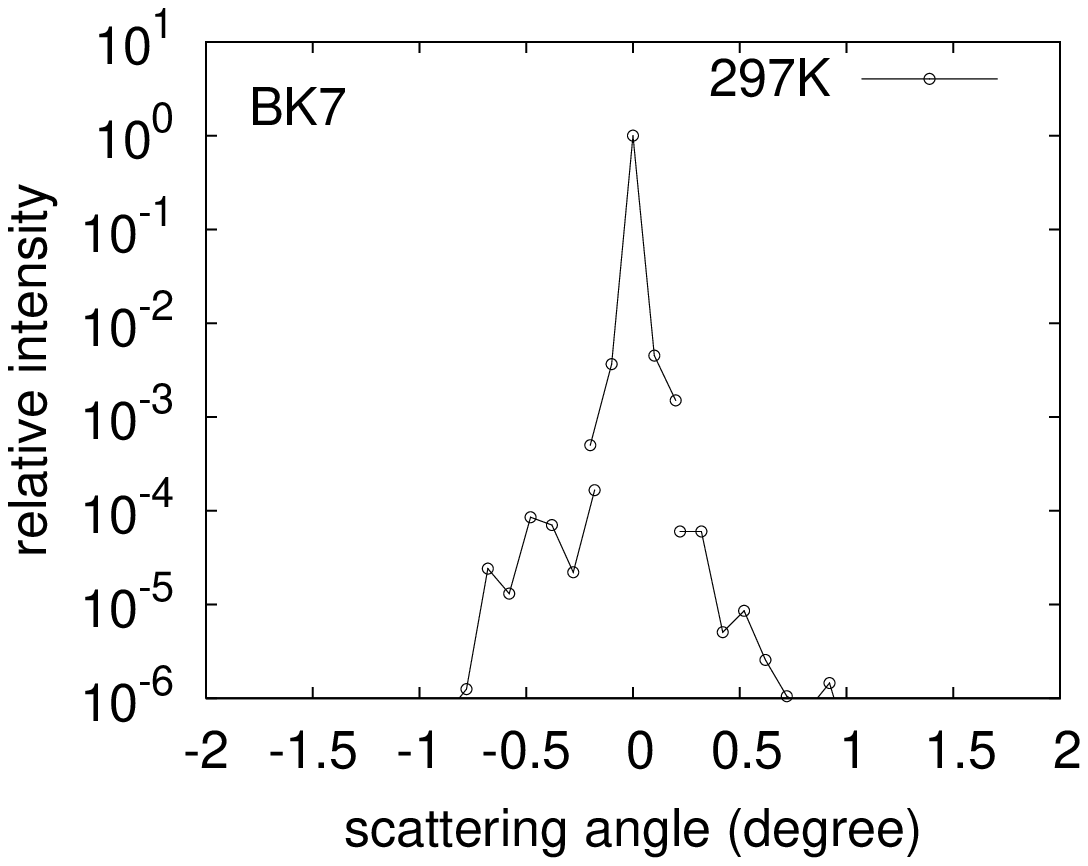}
\caption{Enya et al.}
\label{fig06}
\end{center}
\end{figure}
\end{center}

\vspace{5mm}
\noindent Figure  4.
Top: Microscopic surface structure of the 160\,mm spherical
mirror of the bare C/SiC composite. Left and right figures are the
results of interferometer measurements and photographic images of the
flat part obtained using the WYKO NT1100, respectively. The average
roughness is 20\,nm rms. The observed structures are similar to those
of the flat test mirrors shown in figure {\ref{fig02}}. 
Middle: Left and right
show the spherical part of the mirror by the same method. The average
roughness of this part is 100\,nm rms. 
Bottom: Joint in the 160\,mm
spherical mirror of the bare C/SiC composite. Left and right figures
are the results of interferometer measurements and photographic
images. The average width of the join is 0.5\,mm, where no carbon-fiber
is found. The bright region in the right figure corresponds to
residual silicon. The silicon region is 20-50\,nm lower than the SiC
regions.

\vspace{5mm}
\noindent Figure  5.
Configuration used for the scattered light measurements.

\vspace{5mm}
\noindent Figure  6.
Light scattered from flat test mirrors at room temperature
and cryogenic temperatures. The top and middle figures show the
results of the improved C/SiC bare mirror and the SiSiC coated
mirror. The bottom figure shows the data for a BK7 glass mirror as a
reference.

\newpage

\clearpage

\clearpage

\clearpage

\clearpage

\clearpage

\end{document}